# INTERPRETER – Tool for non-technical losses detection

Hans Bludszuweit
Smart Grids Operation Group Energy Fundación CIRCE
Zaragoza, Spain
0000-0002-3441-7133

Nurseda Y. Yurusen
Analysis & Optimization of Renewable Energy Group
Fundación CIRCE
Zaragoza, Spain
0000-0002-1206-9756

Pablo López Pérez
Smart Grids Manager
Cuerva
Granada, Spain
plopez@cuervaenergia.com

Diego Martínez-López
Smart Grids Operation Group Energy Fundación CIRCE
Zaragoza, Spain
dmartinez@fcirce.es

*Abstract*— This article presents a tool for the detection of non-technical losses, which is being developed within the European INTERPRETER project. The tool employs a hybrid method based on feature detection from smart meter data and grid model analysis. This paper focuses on the grid model analysis, where voltage deviations between the grid model (digital twin) and real-world measurements at a low-voltage pilot site have been evaluated. Energy measurements from smart meters represent hourly mean power, while voltage measurements are instantaneous with uneven time intervals. Thus, measurements are not synchronous, which poses a major challenge for grid analysis. The proposed method focuses on daily mean, minimum, and maximum voltage and results show that deviations in daily minimum voltage are the most useful ones. A heatmap is developed, which helps the DSO expert to have a quick overview of all deviations of all meters in a certain time interval (1-day time step). A total of 6 locations have been identified where field inspections will be done.

*Keywords—electricity theft, INTERPRETER, digital twin, distribution network analysis, voltage analysis*

## I. Introduction

The global electricity demand rose more than 5% in 2021 [5], [10]. Nevertheless, addressing such demand in a cheap and environmentally friendly way was not possible, which resulted in record-breaking highs in electricity market prices and $CO_2$ emissions [6], [8]. In such a challenging market, non-technical energy losses due to fraudulent consumption increase the burden of additional costs for end customers and utility companies. The motivation behind the electricity theft might vary from bitcoin mining [7], cannabis cultivation [1], [4] or just about any other activity. Where smart meters have been deployed, at the end of each day, hourly consumption profiles of each consumer are available in the databases of grid operators. Nevertheless, this is big data and most of the time it is stored in a data lake, which requires careful data management, in order to use it as an input data source for fraudulent behaviour detection. Raw data from smart meters contains numerous errors and requires thorough pre-processing, before being ready for data services, such as NTL detection [13].

The widely accepted detection scheme for non-technical losses (NTL) is to perform on-site inspections by trained expert technicians. These inspections require planning of the visits considering pre-set criteria such as detecting prolonged time intervals without electricity consumption or disproportional demand reduction. The proper definition of the required criteria is of utmost importance in order to avoid costly and futile on-site visits. Low-voltage network visibility is still very low which makes fraud detection very difficult in practice. Introducing grid models in combination with smart meter readings helps to improve the quality of inspection criteria by reducing the uncertainty, where the operators have to look to find the actual fraud location.

In the literature, there exist many published studies in order to perform NTL analysis and fraud detection [2], [3], [12], nevertheless there is still a need for the development of a standardised compact tool. Various studies that the confusion matrix metrics and site visit verified electricity theft detection, but decision-maker interactive candidate list generation visualisation tools are not presented thoroughly. It must be highlighted that site visits are dependent on utility company resources (availability of site technicians) and machine learning models generate long lists of suspicious clients. Therefore, in practice there is still a need for human interaction to fine-tune final site-visit plans. Timely site visits have vital importance since electricity theft might not be continuous. It is also observed that machine-learning models are dependent on site-visit detection dates, which introduces a bias and uncertainty in the model since it is not possible to label all NTL cases in the training data due to the dependency on fraudulent usage verification dates. In this study, by following data science principles, and using anomaly detection visualisation techniques, an impactful and easy to interpret data-driven NTL detection application is presented as a site visit fine-tuning tool and real-time NTL detection technique.

In the European INTERPRETER project [14], a platform-based approach has been developed, in order to bring together heterogeneous information from grid operators, such as grid inventory and smart meter data. A key feature is a grid modelling tool, which provides distribution grid models (digital twins) from available inventory data. With this grid model at hand, smart meter data gain much higher value, as state estimators and load-flow analyses can be undertaken [9]. In the context of this project, an NTL tool is being developed that combines more classical data-driven machine learning with network analysis based on grid models and smart meter data. This hybrid approach is shown in Fig. 1. The machine-learning (ML) component is under development, employing Python Pycaret [15] and R Kernlab [14] libraries, where the Support Vector Machine (SVM) method is showing the best performance in terms of execution speed and forecast indicators (true positives in relation to false positives and false negatives).

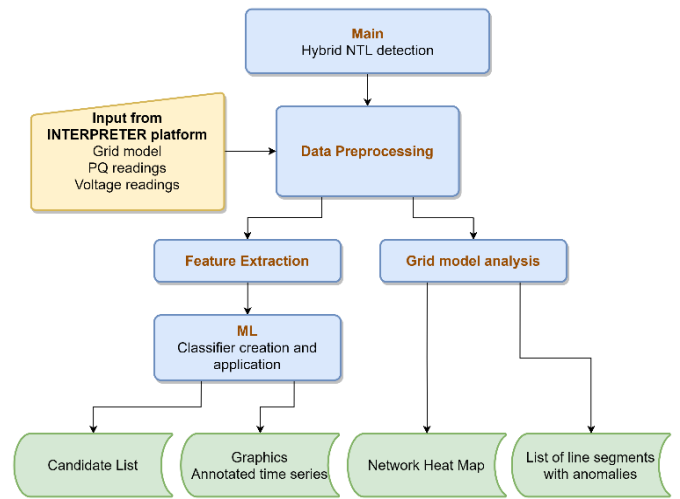

Fig. 1: Flowchart of the INTERPRETER hybrid NTL detection tool.

This paper is focussing on a demo application of the grid model analysis.

The structure of this paper is as follows: Section II describes the INTERPRETER platform where the tool is integrated. The next section gives an overview of considered signals and input data. In section IV methodology is explained and in section V results are presented. The final section summarises main outcomes and discusses the usefulness and limitations of this study.

## II. INTERPRETER PLATFORM

The INTERPRETER platform is a modular grid management solution that provides data services based on an integrated grid modelling tool. Improved LV grid observability, provided by the grid modelling tool, enables advanced monitoring, O&M, planning and flexibility services, saving money and resources. As shown in Fig. 2, the 10 software applications plus the grid modelling tool can be classified regarding their type of service: 5 tools focus on grid operation and maintenance and another 5 on grid planning. The NTL- detection tool is part of the toolset for grid operation and maintenance. More details can be found in D6.1 of this project, published with the title "Integration of data sources in a multiplatform Data Space" on the project website [14].

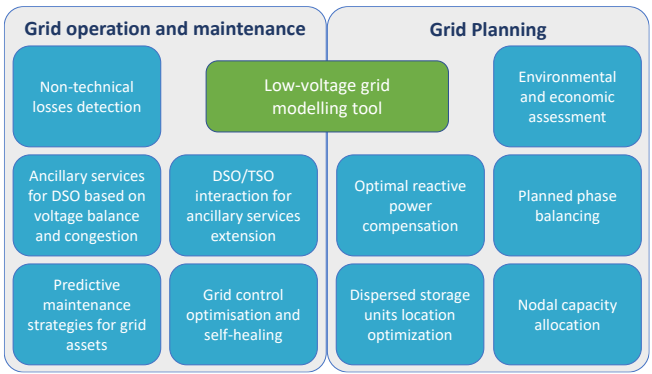

Fig. 2: Classification of INTERPRETER tools regarding application.

### III. DATA

#### A. Spanish Demo-Cuerva

As shown in Fig. 1, input data includes smart meter readings (historical data for voltage, active and reactive power). Within the INTERPRETER project, this kind of data has been provided by the Spanish DSO and project partner Cuerva. The Cuerva Pilot is a 400-V low-voltage (LV) network which consists of 12 feeders serving a total of 266 clients. The grid model contains 690 buses and a variety of loads and line types, including single-phase and 3-phase connections. Fig. 3 shows the network model diagram of the pilot.

The different branches (feeders) that make up the network are shown in different colors. The network model has been created from Cuerva inventory data and is analyzed within DiGSILENT PowerFactory software [16].

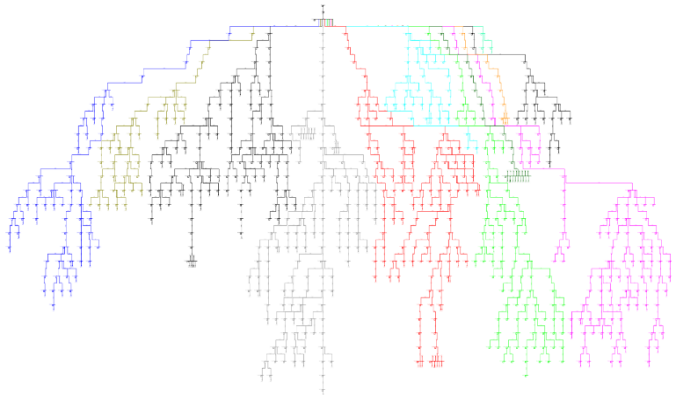

Fig. 3: Topology of the Cuerva pilot network model.

#### B. Smart meter data

Available smart meter readings are hourly energy values, aimed to provide a basis for monthly billing. Therefore, these values represent hourly mean power values instead of instantaneous measurements.

Voltage readings are not necessary for billing purposes, but due to the increased awareness of the usefulness of this kind of data for network monitoring, DSOs are starting to register voltage readings too. In Europe, most smart meters use narrowband powerline communication (PLC) which is sufficient for billing, but results a major challenge which additional information, such as voltage measurements are requested [11].

Within the pilot of the Cuerva network, instantaneous voltage readings have been provided for approximately 11 months, which are the basis of the network analysis presented in this paper. For these instantaneous voltage readings, no synchronized hourly data collection was available. Instead, an automated process sends calls to smart meters one by one. Communication delays are important, which results in voltage readings of approximately one value per hour and meter, but intervals are not constant. Also, missing measurements are very common, as not always the request for voltage measurement is responded to before the next call is sent.

### IV. METHODOLOGY

#### A. Comparison of simulations vs. measurements

As mentioned in the introduction, the Demo application presented here focuses on the network analysis part of the hybrid NTL detection tool. For this purpose, load-flows have been calculated for all available smart-meter power readings, providing hourly voltage values based on average hourly power flow.

On the other hand, instantaneous voltage values are not synchronized, and time intervals vary. This is not an ideal situation, but the aim of the project was to develop a prototype with the available information, which represents a typical DSO with deployed smart meters.

The solution has been to analyze daily mean, minimum and maximum voltages, in order to compare measured and simulated voltages in a meaningful way. The following data treatment procedure was implemented:

- Calculate hourly voltages from smart meter readings
- Calculate daily mean, minimum and maximum voltages from simulation results and measurements
- Calculate the difference between simulated and measured daily values.

In a more formal way, three indicators have been calculated for each meter and each day in the observed time interval of 11 months. The mathematical formulation for this simple calculus is shown in the following equations.

$$\Delta V_{mean} = V_{mean,sim} - V_{mean,meas}$$
$$\Delta V_{min} = V_{min,sim} - V_{min,meas}$$
$$\Delta V_{max} = V_{max,sim} - V_{max,meas}$$

where $\Delta V_{mean}$, $\Delta V_{min}$, $\Delta V_{max}$ are the three daily indicators for deviation between simulation and measurements. The sub-indices "sim" and "meas" stand for simulated and measured values.

*B. Statistical analysis of voltage deviations*

In Fig. 4, histograms are shown for differences of daily minimum, maximum and average voltages, observed in measurements and simulation. The big picture shows that almost all deviations are within the range of ±0.1 p.u., which are reasonable results, considering that:

- There are some meters with wrong phase information.
- There are three-phase meters that are actually single-phase and vice versa.
- The real network has more meters than there are in the model. (The model is not up to date due to a lack of information considering location of the new customers.)

A second important conclusion from Fig. 4 is that voltage deviations are quite different, depending on the indicator which is chosen. Calculating the statistical parameters of average and standard deviation, the differences between these three indicators can be quantified, as shown in TABLE I. The smallest deviations are observed for daily average and maximum voltage ($\Delta V_{mean}$, $\Delta V_{max}$), while daily minimum voltage ($\Delta V_{min}$) shows the largest deviations. In addition, a remarkable positive bias indicates that simulated voltages tend to be higher than measured ones. This is coherent with the fact, that simulations are based on hourly mean loads.

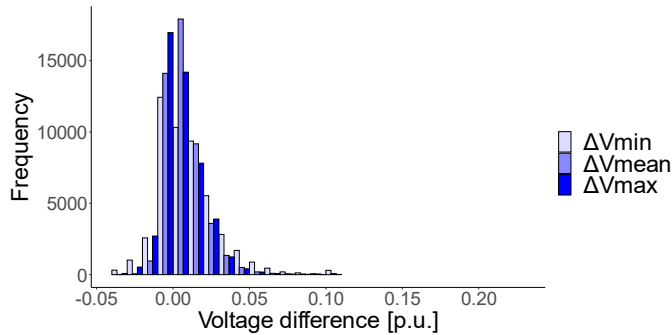

Fig. 4: Histograms of min, mean and max voltage difference.

TABLE I. STATISTICAL PARAMETERS OF MIN, MEAN AND MAX VOLTAGE DIFFERENCE.

|  | **ΔVmean** | **ΔVmax** | **ΔVmin** |
|---|---|---|---|
| Average | 0.0076 | 0.0054 | 0.0112 |
| Standard deviation | 0.0128 | 0.0128 | 0.0207 |

This observation already indicates that $\Delta V_{min}$ is the most valuable of the three indicators. Daily maximum voltage is typically observed during low-load periods, where it is hard to identify any anomalies. Daily average voltage may not be selective enough, as periods of low and high demand are mixed. Finally, daily voltage minima indicate demand peaks, and thus, carry valuable information regarding anomalies.

V. RESULTS

Results have been evaluated for all three indicators, but the heatmaps from $\Delta V_{mean}$, $\Delta V_{max}$ do not provide additional useful information, as will be explained later. Therefore, in this section, results of the voltage analysis are presented only for indicator $\Delta V_{min}$.

The heat map in Fig. 6 shows voltage deviations between simulation and field measurements for all meters over time. Each time step is one daily value. As mentioned before, the daily time step was chosen, in order to reduce the important noise from the non-synchronized input data.

In Fig. 6 the heatmap of $\Delta V_{min}$ (difference between simulation and field measurements for daily minimum voltages) is shown. Negative values (green shades) indicate, that measured daily minimum voltage was higher than the simulated one. This occurs if actual demand peaks are lower than expected (simulated) ones or distributed generation is not being captured. Considering that simulations are based on hourly mean demand and field measurements are instantaneous, typically positive values of $\Delta V_{min}$ are expected and negative values should be rare.

In this sense in Fig. 6 the oldest data can be considered as systematically erroneous. During May 2021, a major data gap indicates some modifications in the data collection process for instantaneous voltage measurements. After this gap, results appear to be much more coherent with what could be expected.

Another interesting observation is, that there are meters with large deviations (> 0.1 p.u.) only until a certain date, and others with deviations which start at some point.

Heatmaps have also been created for $\Delta V_{mean}$ and $\Delta V_{max}$. For $\Delta V_{mean}$, very similar patterns have been observed, but with less contrast (smaller deviations in general). For $\Delta V_{max}$, compared to the picture obtained for $\Delta V_{mean}$ and $\Delta V_{min}$, most of the anomalies have disappeared. Also, negative values are more frequent in the time interval after the gap of measurements. Although statistical parameters such as average and standard deviation were very similar to $\Delta V_{mean}$, the heatmap is very different. Thus, it is concluded that $\Delta V_{max}$ is the least useful indicator to identify anomalies.

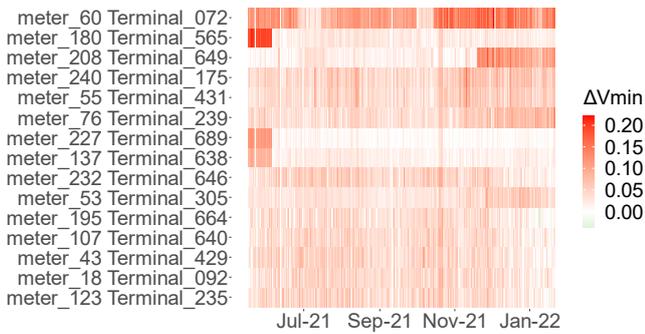

Fig. 5: Heatmap of top 15 locations with highest values of ΔVmin_max.

limited number of meter locations, where significant deviations patterns are observed. In order to provide a further step of data filtering, a criterion has been established in order to identify the most critical cases. This criterion has been chosen to be the maximum value ΔVmin_max in the observed time interval, as it captures even short but large deviations.

In Fig. 5 a reduced heatmap is shown, including only the 15 locations with the highest values of ΔVmin_max, sorted from higher to lower values from top to bottom.

In addition, the first period of the time series, where inconsistent data was detected, has been removed, to obtain a clearer picture.

The heatmap in Fig. 6 is already a good overview for the expert team for NTL detection of a DSO, but there is only a

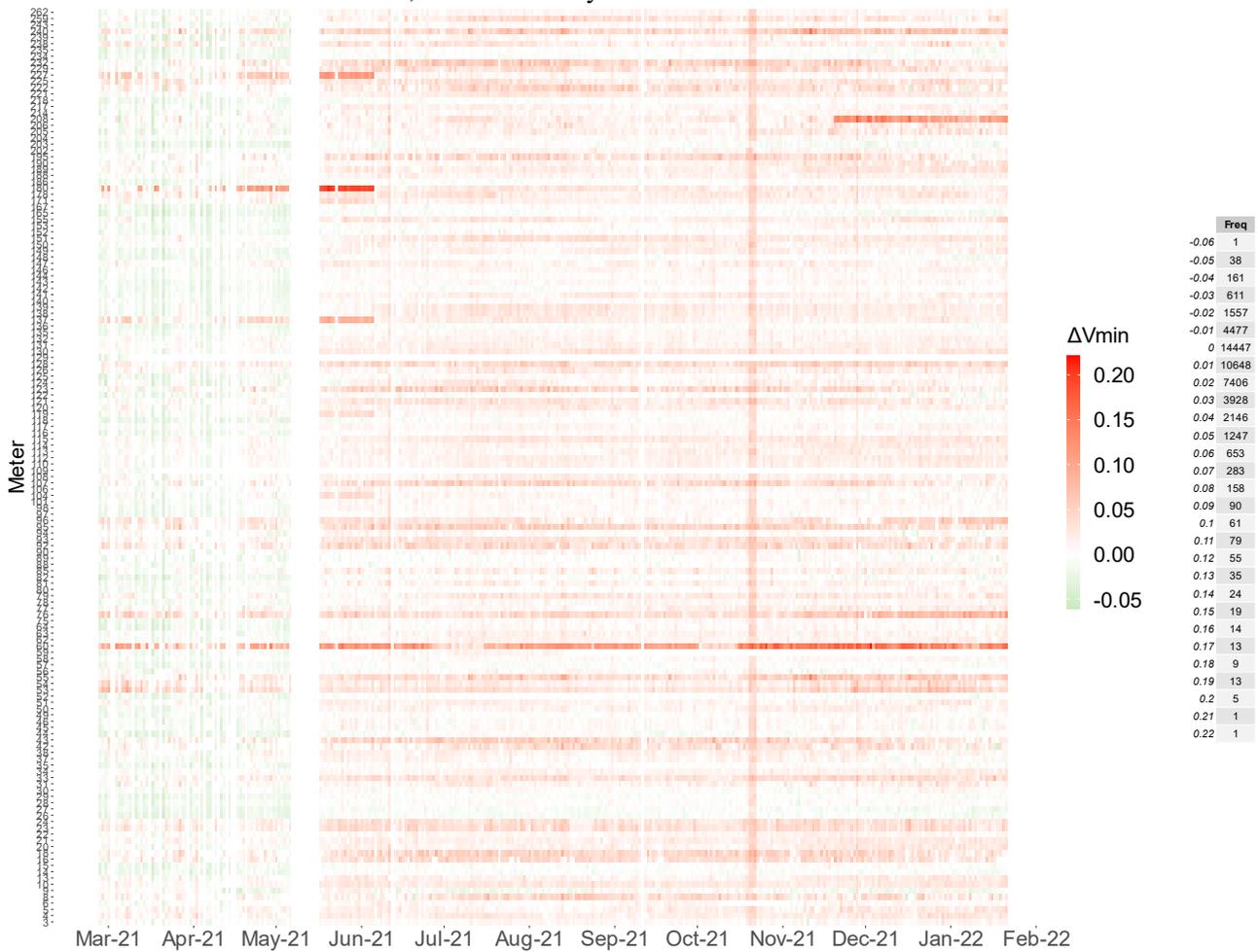

Fig. 6: Heatmap with difference of daily minimum voltage ΔVmin over time.

TABLE II. TOP 15 LOCATIONS WITH HIGHEST VALUES OF ΔVmin_max.

| Rank | Meter_ID | Bus | ΔVmin_mean | ΔVmin_max | Comments |
|---|---|---|---|---|---|
| 1 | meter_60 | Terminal_072 | 0.0951 | 0.2211 | Field inspection candidate |
| 2 | meter_180 | Terminal_565 | 0.0361 | 0.2120 | Candidate for validation |
| 3 | meter_208 | Terminal_649 | 0.0316 | 0.1834 | Field inspection candidate |
| 4 | meter_240 | Terminal_175 | 0.0442 | 0.1588 | Field inspection candidate |
| 5 | meter_55 | Terminal_431 | 0.0374 | 0.1457 | Field inspection candidate |
| 6 | meter_76 | Terminal_239 | 0.0416 | 0.1274 | Field inspection candidate |
| 7 | meter_227 | Terminal_689 | 0.0168 | 0.1254 | Candidate for validation |
| 8 | meter_137 | Terminal_638 | 0.0217 | 0.1038 | Candidate for validation |
| 9 | meter_232 | Terminal_646 | 0.0342 | 0.1020 | Discard (low recent deviations) |
| 10 | meter_53 | Terminal_305 | 0.0341 | 0.1020 | Field inspection candidate |
| 11 | meter_195 | Terminal_664 | 0.0266 | 0.1016 | Discard |
| 12 | meter_107 | Terminal_640 | 0.0296 | 0.0976 | Discard |
| 13 | meter_43 | Terminal_429 | 0.0325 | 0.0970 | Discard |
| 14 | meter_18 | Terminal_092 | 0.0323 | 0.0959 | Discard |
| 15 | meter_123 | Terminal_235 | 0.0319 | 0.0949 | Discard |

In this reduced heatmap, 5 meters show interesting patterns, which may indicate anomalies that are not related to the grid model:

- **Meter 60**: Intermittant deviations, with 2 low-deviation periods in July and October 2021 → this is a candidate for closer analysis and field inspection.
- **Meters 180, 227 and 137**: deviations cease early June 2021, which may indicate some field intervention → They are candidates for validation of the method
- **Meter 208**: Deviations start within November 2021 and last till the end of the observation interval. This is another candidate for closer analysis and field inspection.

The remaining locations should be examined by DSO experts, in order to establish the final list for field inspection. Those with lower deviations towards the end of the observed interval might be removed, which results in a reduced list of 6 meters (60, 208, 240, 55, 76 and 53), which are all ranking within the top 10 of the list.

In TABLE II. the second output of the grid model analysis is shown. It contains the list of line segments (terminals), represented by terminal IDs where the highest values of ΔVmin_max have been observed. This list is sorted in the same manner as the heatmap in Fig. 5. For information purposes, apart from indicator ΔVmin_max, also ΔVmin_mean (average over the observed time interval) has been included. With this list, DSO operators can check directly if identified anomalies have been solved already, or if field inspections are indicated.

VI. CONCLUSION

Massive roll-out of smart metering in distribution grids provides a large amount of information for NTL detection. Non-synchronous measurements are a serious challenge for network analysis. Nevertheless, most of the existing measuring infrastructure in the distribution grid is still far away from synchronized measurements. Therefore, methodologies are required to deal with this condition.

The results presented in this paper are encouraging, as relevant voltage deviations between simulation and real-world measurements could be visualized, despite the large amount of uncertainty. It becomes clear that small frauds will not be detected, due to the noise in the data. But large deviations and modifications in the grid have become clearly visible. In fact, the visualization procedure is very important and has not been addressed in other work in the literature, to the extent as presented here.

Non-technical losses include both, fraud and not measured consumption. Therefore, this methodology is a very helpful tool to indicate the location and severity of anomalies in the network, even if the model is not perfect, as it was the case here.

The INTERPRETER project is still on-going and field verification has just started. The very limited number of fraud cases detected manually within the observed time interval could not be confirmed with the voltage analysis, as volumes were small. Nevertheless, detected hot spots will be revised and results will serve to improve the grid model in an iterative process.

Future work includes the combination of the ML model with the grid model in order to create a more comprehensive list of candidates. On the other hand, as this voltage analysis is able to identify locations with large deviations, it will be integrated into the grid modelling tool, which has been mentioned in the introduction.

ACKNOWLEDGMENTS

The INTERPRETER project has received funding from the European Union's Horizon 2020 Innovation Action programme under grant agreement No 864360.